\documentclass[12pt]{article}

\usepackage{amssymb}

\def\be{\begin{equation}}
\def\ee{\end{equation}}
\def\bea{\begin{eqnarray}}
\def\eea{\end{eqnarray}}

\def\p{\partial}
\def\a{\alpha}

\def\vfi{\varphi}

\def\wt{\widetilde}

\def\cN{{\cal N}}

\def\Im{{\rm Im }\,}
\def\Re{{\rm Re }\,}

\def\pt{{$\cal PT$}}

\newcommand{\Sc}{Schr\"odinger }

\newtheorem{Prop}{Proposition}

\begin{document}

\title {\large\bf Irreducible second order SUSY transformations\\
between real and complex potentials}

\author{
  Boris F. Samsonov
}

\date{\small Department of Physics, Tomsk State
 University, 36 Lenin Avenue, 634050 Tomsk, Russia}

\maketitle

\begin{abstract}
Second order SUSY transformations between real and complex
potentials for three important from physical point of view
Sturm-Liouville problems, namely, problems
with the Dirichlet boundary conditions for a finite interval, for
a half axis and for the whole real line are analyzed.
For every problem conditions on transformation functions are
formulated when transformations are irreducible.

\end{abstract}

%%%%%%%%%%%%%%%%%%%%%%%%%%%%%%%%%%%%%%%%%%%%%%%%%%%%%%%%%%%%%%%%%%%%%%%%%
%%%%%%%%%%%%%%%%%%%%%%%%%% INTRODUCTION %%%%%%%%%%%%%%%%%%%%%%%%%%%%%%%%%

\section{Introduction}

Non-Hermitian operators started to
attract attention of physicists
soon after the main foundation of quantum theory was built
(see e.g. \cite{Pauli,Gupta}).
A recent numerical observation
\cite{Bender}
that some non-Hermitian one-dimensional
Hamiltonians may have purely real spectrum
 re-initiated an attempt to
generalize quantum mechanics by accepting non-Hermitian operators
for describing physical observables \cite{BBJ}
(so called `complex quantum mechanics',
for recent developments see e.g. \cite{BCM,ChJP}).
In this respect it is worthy of special mention
the paper \cite{QuasiH}
where the authors
establish a general criterion for a set of
non-Hermitian operators
(so called quasi-Hermitian)
to constitute a consistent quantum mechanical system with a normal
quantum mechanical interpretation.

Another class of non-Hermitian operators, called pseudo-Hermitian,
was introduced
 by Dirac and Pauli
 and later used by
Wick and Lee \cite{Pauli} to overcome some difficulties
related with using
Hilbert spaces with an indefinite metric.
Recently due to Solombrino and Scolarici \cite{SS}
 this concept found a further
 generalization  as weak pseudo-Hermiticity.
A recent observation \cite{Mostafa}
 that the real character of
the spectrum of a pseudo-Hermitian
 Hamiltonian $h$ is related to the existence of a
pseudo-canonical transformation,
 which makes $h$ similar to a Hermitian operator,
permits us to suppose that there exists a certain overlap between
these two classes of non-Hermitian operators,
 appearing to be the
most appropriate candidates for describing physical observables.

From the general point of view if some non-Hermitian
operators with a purely real spectrum are similar to Hermitian
ones their incorporation into quantum mechanics cannot be
considered as a more general approach with respect to the
conventional quantum mechanical description.
Their use may give
(or may not give, for a recent discussion see \cite{MostafaP})
only calculational advantages \cite{BCM,QuasiH}.
From the first glance this observation leads to the
negative answer to the question
whether
or not complex quantum mechanics is an extension of the
conventional quantum mechanics.
But if we take into account the fact that between non-Hermitian
operators with a purely real spectrum there exist such operators
which never can be similar to Hermitian ones the above question
seems to be still open.
In particular, a non-diagonalizable
operator has an incomplete system of eigenfunctions and therefore
it can never be similar to a Hermitian operator, the eigenfunctions
of which form a complete basis in corresponding Hilbert space.
Other operators which should be studied in this respect are
the ones having spectral singularities in the continuous part of
the spectrum, the feature which never appears in the Hermitian case
\cite{Naimark}.

 In \cite{Z2} an overlap between
 \pt-symmetric quantum mechanics (by this term some authors mean
 a  complex extension of quantum mechanics)
 and supersymmetric quantum mechanics
 (SUSY QM, for recent reviews see  \cite{JPA}) was noticed.
Furthermore, as it was shown
SUSY QM may be useful to transform non-diagonalizable
Hamiltonians into diagonalizable forms \cite{MyLett1}
and remove spectral singularities from their continuous spectrum
 \cite{MyLett2}.
These results permit us to hope that SUSY QM may be a very
useful tool in complex quantum mechanics.

In this letter we are using the term {\em SUSY transformations}
in its narrow sense
as differential transformations between two (exactly
solvable) Hamiltonians having almost the same
(up to a finite number of levels) spectra and do not discuss
underlying algebraic constructions.

Supersymmetric transformations involving first order intertwining
operators
between one-dimen\-si\-on\-al Hamiltonians
\be\label{h01}
h_{0,1}=-\p_x^2+V_{0,1}(x)\,,\qquad \p_x\equiv d/dx\,,\qquad
x\in(a,b)
\ee
with the possibility for the potentials $V_{0,1}(x)$ to be
complex-valued functions
were studied in \cite{Cannata1,Cannata2}.
A succession of two SUSY transformations with equal factorization
constants (confluent transformations, see e.g. \cite{revs})
 was used in \cite{Milanovich}
 to obtain bound states embedded into continuum
of scattering states of a complex potential.
On the other hand it is clear that if the intermediate Hamiltonian
$\wt h$ of a chain of two
first order
 transformations $h_0\to\wt h\to h_1$ has
a physical meaning there are no special needs to study the
second order
transformation leading from $h_0$ to $h_1$ directly
(2-SUSY transformation); all
properties of the Hamiltonian $h_1$ can be understood at the level
of the first order transformation $\wt h\to h_1$ (1-SUSY transformation).
This is not the case if $\wt h=-\p_x^2+\wt V(x)$
 is not a well defined Hamiltonian
acting in the same Hilbert space as $h_0$ and $h_1$.
In this respect for the case when
both $V_0(x)$ and $V_1(x)$ are real
 in \cite{Andr2} the notion of reducible and
irreducible SUSY transformations was introduced.
The chain is
called reducible if $\wt V(x)$ is real \cite{Andr2} and irreducible
otherwise.
Evidently as far as complex potentials are concerned
chains irreducible in this sense become reducible \cite{Cannata1}.
Later \cite{S}
(for a recent discussion see \cite{AndrCan})
another type of irreducible transformations was described. They
appear when the potential $\wt V(x)$ has singularities inside the
interval $(a,b)$ where the potentials $V_0(x)$ and $V_1(x)$ are regular.
Recently the third possibility for
irreducible chains was noticed \cite{SP}.
It appears if the intermediate potential $\wt V(x)$
is regular inside $(a,b)$ but the spectrum of $\wt h$ is
completely different of the spectrum of $h_0$ and
 no SUSY transformations between their eigenfunctions exist
 whereas $h_0$ and $h_1$ are (almost) isospectral and their
 eigenfunctions are connected with the help of a second order SUSY
 transformation.
 For these two kinds of irreducible chains spectral
 properties of $h_1$ cannot be derived from spectral properties of
 $\wt h$
even in the case of complex potentials
 and one needs to analyze second order SUSY
 transformations between $h_0$ and $h_1$
 without involving the intermediate potential
 $\wt V(x)$.
In this Letter we formulate conditions for 2-SUSY transformations
between a real potential $V_0$ and a complex potential $V_1$
with a purely real spectrum
to be irreducible.

\section{First and second order SUSY transformations}

In this section we briefly review the main properties
of first and second order SUSY transformations
(for details see e.g. \cite{JPA,revs,BS})
we need below.

We consider second order ordinary differential equations
(\Sc equations)
\be\label{1}
(h-E)f_E(x)=0\qquad x\in (a,b)
\ee
with (Hamiltonian)
 $h=h_0,\wt h, h_1$ and $f_E(x)=\psi_E(x),\wt\psi_E(x),\vfi_E(x)$
respectively; $E\in \Bbb C$ is a parameter and $a$, $b$
 may be both finite and infinite.

We say that the Hamiltonians
$h_0$ and $\wt h$ are 1-SUSY partners if there exists a first
order differential operator $\wt L_1$ intertwining $h_0$ and
$\wt h$, $\wt L_1h_0=\wt h\wt L_1$.
Similarly, $\wt h$ and $h_1$ are 1-SUSY partners if there exists
a first order differential operator
$L_1$ such that $L_1\wt h=h_1L_1$.
Evidently, the second order differential operator $L=L_1\wt L_1$
intertwines $h_0$ and $h_1$
\be\label{inter}
 Lh_0=h_1L
 \ee
 and $h_1$ is called 2-SUSY partner for $h_0$.
 Once the existence of $L$ is established
  solutions of equation (\ref{1}) with $h=h_1$
 can be found by applying $L$ to solutions
 of the same equation with $h=h_0$,
$
 \vfi_E=L\psi_E$, $ \psi_E\notin\mbox{ker}L$.
 Evidently, similar property takes place for
 solutions $\wt \psi_E$ of the \Sc equation with
 the intermediate
 Hamiltonian $\wt h$;
they are expressed in terms of solutions of the initial
equation $\psi_E$
and $u_1(x)$, $h_0u_1(x)=\a_1 u_1(x)$, $\a_1\in\Bbb C$
($u_1$ is called transformation function and $\a_1$ is
factorization constant)
\be\label{1orderPsi}
 \wt\psi_E(x)=\wt L_1\psi_E(x)=
 -\psi'_E(x)+w(x)\psi_E(x)\qquad w(x)=u_1'(x)/u_1(x)\qquad
 E\ne\a_1
 \ee
 \[
\wt\psi_{\a_1}(x)=\frac{1}{u_1(x)}\,.
 \]
 The potential $\wt V(x)$ is expressed in
terms of the function $w(x)$ as follows:
\be\label{1orderV}
\wt V(x)=V_0(x)-2w'(x)\,.
\ee

For the next transformation step $\wt h\to h_1$ the same formulas
(\ref{1orderPsi}) and (\ref{1orderV})
 with the evident modifications
should be used with the only difference that now one distinguishes
the confluent case,
when the factorization constant $\a_2$ at the second step
of transformation coincides with
that of
the first step,
 $\a_2=\a_1$, from the usual
(non-confluent) case when these constants are
different $a_2\ne\a_1$.

Using the second order transformation operator
$L=L_1\wt L_1$ one can avoid
the intermediate step and go from $h_0$ to $h_1$ directly
\be\label{V1}
V_1=V_0-2 \left[\log W(u_1,u_2)\right]''
\ee
\be\label{fiE}
\vfi_E=L\psi_E=W(u_1,u_2,\psi_E)/W(u_1,u_2)\,.
\ee
Here and in the following the symbol $W$ with arguments being
functions
denotes Wronskians, $u_1$,
$u_2$ and $\psi_E$ are solutions to equation (\ref{1})
with $h=h_0$
corresponding to the eigenvalues $\a_1$, $\a_2$ (factorization
constants),
%\be\label{h0a}
$h_0u_{1,2}=\a_{1,2}u_{1,2}$,
%\ee
 and $E$ respectively.
Expressions (\ref{V1}) and (\ref{fiE}) are known as
particular cases of
Crum-Krein formulas
\cite{Krein}.

Formula (\ref{fiE}) defines the operator $L$ for any sufficiently
smooth function  $\psi_E$ but if $\psi_E$ is a solution to equation
(\ref{1}) other forms of this equation are useful
\bea\label{fi1}
&\vfi_E &
=(E-\a_2)\psi_E+(\a_1-\a_2)
\frac{W(u_2,\psi_E)}{W(u_1,u_2)}\,u_1\\
& &
=(E-\a_1)\psi_E+(\a_1-\a_2)
\frac{W(u_1,\psi_E)}{W(u_1,u_2)}\,u_2\,. \label{fi2}
\eea
Here the use of equation (\ref{1}) has been made
to express the second derivatives of the functions $u_{1}$,
$u_{2}$ and
$\psi_E$ in terms of the functions themselves.
Operator $L$ as given in (\ref{fi1}-\ref{fi2})
maps any two-dimensional space of solutions of equation (\ref{1})
with $h=h_0$ at $E\ne\a_1,\a_2$
onto corresponding space of solutions of the same equation with
$h=h_1$.
The two-dimensional space $\mbox{span}(u_1,u_2)$ is the kernel
of $L$, $Lu_{1,2}=0$.
Despite that with the help of $L$ one can find
solutions of the transformed equation corresponding to $E=\a_1,\a_2$.
For this purpose one has to act with $L$ on functions
$
v_{1,2}\ne u_{1,2}$,
$ h_0v_{1,2}=\a_{1,2}v_{1,2}$.
 Using the fact that $W(u_{1,2},v_{1,2})=\mbox{const}$
and putting $\psi_E=v_{1,2}$, $E=\a_{1,2}$
 in (\ref{fi1}) and (\ref{fi2})
 one readily gets
\be\label{fial}
\vfi_{\a_{1,2}}=\frac{u_{2,1}}{W(u_1,u_2)}\qquad
h_1\vfi_{\a_{1,2}}=\a_{1,2}\vfi_{\a_{1,2}}
\ee
where we have omitted an inessential constant factor.
It is worth to note that the use of these functions for the next
step of the second order  transformation gives back the initial
Hamiltonian $h_0$ and, hence, the procedure is completely
reversible.
Our last comment here is that
as it follows from (\ref{V1})
 to obtain nonsingular for $x\in (a,b)$ potential differences
 it is necessary
 that $W(u_1,u_2)\ne 0$ $\forall x\in(a,b)$
  which will be supposed to be the case.

 For equal factorization constants
 ($\a_1=\a_2=\a$, the confluent case) the function
 $W(u_1,u_2)$ in (\ref{V1}) should be replaced by
\be\label{WConfl}
 W_c(x)=c+\int_{x_0}^xu^2(y)dy\,.
\ee
Constants $c$ and $x_0$ should be chosen such that $W_c(x)\ne 0$
$\forall x\in (a,b)$.
Solution of the \Sc equation with the Hamiltonian $h_1$ are given
by
\be\label{fiConfl}
\vfi_E(x)=L\psi_E=\left( \a-E \right)\psi_E(x) +
\frac{W(\psi_E,u)}{W_c(x)}\,u(x)\,.
\ee
This formula gives a solution for $E=\a$  also provided
$\psi_\a$ is linearly independent with $u$
\be\label{fiConfl}
\vfi_\a(x)=\frac{u(x)}{W_c(x)}
\ee
where  once again we have omitted an inessential constant factor.

 The properties described above take place
irrespective of any boundary value problem related to the
differential equation (\ref{1}).
Here we shall
consider boundary value problems of three kinds
 which are the most interesting
from physical viewpoint:\vspace{-.5em}
\begin{description}
\item[({i})] Regular Sturm-Liouville problem;
the potential $V_0$ is bounded and
 continuous in $[a,b]$ which is a finite interval.
 We will consider only Dirichlet boundary conditions,
$\psi_E(a)=0$, $\psi_E(b)=0$
 imposed on smooth
(infinitely differentiable)
 functions from $L^2(a,b)$
 which form an initial domain of
 definition of $h_0$ which
 has a purely discrete spectrum.
\item[(\bf ii)] Scattering potentials on a semiaxis, i.e. $V_0(x)$ is
 continuous and bounded from below for $x\in[0,\infty)$
 and such that
 \be\label{cond1}
\int_0^\infty x|V_0(x)|dx<\infty\,.
 \ee
 Here we impose on solutions to equation (\ref{1})
the Dirichlet boundary condition at the origin only,
$\psi_E(0)=0$,
which together with the condition of square integrability over the
interval $[0,\infty)$ selects the bound states.
The scattering states have an
oscillating asymptotical behavior.
The initial domain of definition of
 $h_0$ consists of
 infinitely differentiable
functions vanishing for sufficiently large $x$
and at the origin.
The operator $h_0$
 has a finite number of discrete
 levels and the continuous spectrum filling the
 positive semiaxis.
 \item[(iii)] Confining  and scattering potentials
 on the whole real line, $(a,b)=\Bbb R$.
For confining potentials $V_0$
is locally bounded and $V_0(x)\to\infty$ when $|x|\to\infty$.
Scattering potentials are selected by the condition
\be
\int_{-\infty}^\infty |xV_0(x)|dx<\infty\,.
 \ee
 where $V_0(x)$
 is continuous and semi-bounded from below.
Operator $h_0$ is
initially defined on the set of infinitely differentiable
functions from $L^2({\Bbb R})$ vanishing for sufficiently
large $|x|$.
For confining potentials the spectrum is purely discrete.
The scattering potentials have a finite number of
discrete levels and a two-fold degenerate continuous spectrum
filling the positive semiaxis.
\end{description}

In all cases
the operator $h_0$ is essentially self-adjoint  and
 has
a complete set of eigenfunctions
(in the sense of generalized functions for continuous
spectrum eigenfunctions) in corresponding Hilbert space
(see e.g. \cite{Berezin}).

\section{Complex SUSY partners of real potentials}

\subsection{General remarks}

As we shall see both the analysis and the results strongly depend on
the character of the initial Sturm-Liouville problem.
Nevertheless, there is a property common to all eigenvalue
problems which is essential for our analysis.
We formulate it as the following
\begin{Prop}\hspace{-.5em}.
For a real
potential $V_0(x)$
defining a self-adjoint operator $h_0$ in the space
$L^2(a,b)$
any solution $\psi_\a(x)$ to the \Sc equation
with $\Im(\a)\ne0$
 can vanish at no more then in
one point of the interval $[a,b]$.
For nonfinite values of $a,b$
the statement should be understood in the sense of limit, i.e.
for instance for $b=\infty$ this is
$\lim _{x\to\infty}\psi_\a(x)=0$.
If $\Im(\a)=0$ and exists $x_0\in[a,b]$ such that $\psi_\a(x_0)=0$
then up to an inessential constant factor $\psi_\a(x)$ is real
$\forall x\in[a,b]$.
\end{Prop}

Proof.
The first part of the statement follows from the property of a
self-adjoint operator to have only real eigenvalues.
Indeed, if equations
$\psi_{\a}(x_0)=\psi_{\a}(x_1)=0$
 took place
for
$\mbox{Im}(\a)\ne0$ and
$x_0,x_1\in[a,b]$,
 then the
operator $h_{\a}$, defined by the same differential expression
$h_0$ and the zero boundary conditions at the ends of the interval
$[x_0,x_1]$,
 being self adjoint
 would have the complex eigenvalue $E=\a$ which is impossible.

 The second part follows from the property that for a real $\a$
the basis functions, $\psi_{1\a}(x)$ and $\psi_{2\a}(x)$
  in
 the 2-dimensional space of solutions of equation (\ref{1}) can
 always be chosen real so that any complex-valued solution
 $\psi_\a(x)$ is
 a linear combination
 $\psi_\a(x)=c_1\psi_{1\a}(x)+c_2\psi_{2\a}(x)$.
For a finite value of $x_0$
 from the equation
 $\psi_\a(x_0)=0$ one of the constants, say $c_2$
 (if $\psi_{2\a}(x_0)\ne0$)
can be found.
Evidently, it
  is proportional to $c_1$ with a
 real proportionality coefficient.
 If $\psi_\a(\pm\infty)=0$ the function $\psi_\a(x)$ is
 real-valued
 for a real $\a$ up to an inessential constant factor.
The statement follows from a contradiction
which appears
if one supposes the
opposite statement to be true.
Indeed, if $\psi_\a(x)$ is a complex-valued function
then $\psi^*_\a(x)$
(asterisk means complex conjugation)
is linearly independent with $\psi_\a(x)$ and
 $\psi^*_\a(\pm\infty)=0$ which is impossible.
 \qquad $\square$

We impose on solutions to equation ({\ref{1})
 the same zero (Dirichlet) boundary conditions
 after 2-SUSY transformation.
  Thus we have
two boundary value problems, initial and transformed,
which we will denote (I) and (II) respectively.

 For the usual (non-confluent) case, 2-SUSY  transformation is
 reducible if 1-SUSY transformation with either $u=u_1$ or
 $u=u_2$ is `good'.
This is due to the fact that the chain can start from either
$u=u_1$ or $u=u_2$.
This contrasts with the confluent case
since the chain starts now always from 1-SUSY transformation  based on
the transformation function $u$.
Therefore
any 2-SUSY
 transformation is irreducible if this transformation is `bad'.

Below we analyze conditions for the
transformation functions $u_{1,2}$ and factorization constants
$\a_{1,2}$ giving (according to (\ref{V1})) for a given real $V_0(x)$
a complex potential function $V_1(x)$ such that the operator $h_1$
defined in the corresponding Hilbert space according to the cases
(i)-(iii) has a real spectrum
coinciding with the spectrum of
 $h_0$ with the possible
exception of one or two levels
and the transformation is
irreducible.
We illustrate every possibility with the simplest example considering
boundary value problems for $V_0(x)=0$.

\subsection{Regular Sturm-Liouville problem}

 The
problem (I) is regular but the problem (II) may become singular if
the potential $V_1(x)$ is unbounded in one of the bounds
of the (finite) interval $[a,b]$
 or in the both.

To distinguish irreducible second order
transformations from reducible ones we have to start the analysis from
 first order
transformations.
From (\ref{1orderPsi}) it follows that if $\psi_E(a)=0$
 then $\wt\psi_E(a)=0$
 if and only if $u_1(a)=0$.
Hence
to keep the zero boundary conditions we have to choose
the function $u_1(x)$
 vanishing both at $x=a$ and at $x=b$.
 This means that it is
  an eigenfunction of the
problem (I).
Since any eigenfunction except the ground state function
has zeros inside the interval $(a,b)$ we conclude that in this case there
exists the only admissible first order SUSY transformation. It
corresponds to $u_1(x)=\psi_0(x)$ (the ground state function)
after which the ground state
level is deleted.
This means that
any
2-SUSY transformation which does
not involve the ground state function of the problem (I) is
irreducible.

It is clear from  (\ref{fi1}-\ref{fi2})
 that if both $u_1$ (or equivalently $u_2$) and $\psi_E$,
$E\ne\a_1,\a_2$,
 satisfy the zero boundary
conditions then $\vfi_E$
satisfies the zero boundary conditions also. Hence,
to keep the zero boundary conditions after 2-SUSY transformation
we have the
following possible choices for $u_1$ and $u_2$:\vspace{-.5em}
\begin{description}
\item[(a)] $u_1(a)=u_1(b)=u_2(a)=u_2(b)=0$;%
\item[(b)] $u_1(a)=u_1(b)=0$, $u_2(a)=0$ (or $u_2(b)=0$),
$u_2(b)\ne0$ (or $u_2(a)\ne0$);
\item[(c)] $u_1(a)=u_1(b)=0$,
$u_2(a)\ne0$, $u_2(b)\ne0$;
\item[(d)] $u_1(a)=u_2(b)=0$ (or  $u_1(b)=u_2(a)=0)$,
$u_1(b)\ne0$, $u_2(a)\ne0$
(or $u_1(a)\ne0$, $u_2(b)\ne0$).
\end{description}

In case {\bf (a)} both $u_1$ and $u_2$ are eigenfunctions of the
problem (I) and there is no way to get a complex potential
difference.

In case {\bf (b)} $u_1$ is an eigenfunction of $h_0$,
$u_1(x)=\psi_k(x)$ and
$\a_1=E_k$, $k=0,1,\ldots$.
Hence, the level $E_k$ is not present in the spectrum
of $h_1$. The parameter $\a_2$ can take any complex value except
for $\a_2=E_l$, $l=1,2\ldots$, $l\ne k$
 since in this case $u_2=\psi_l$ and we are back in
the conditions of
the case (a).
According to (\ref{fial}) the function $\vfi_{\a_2}$
satisfies the zero boundary conditions and, hence, the point $E=\a_2$
belongs to the discrete spectrum of $h_1$.
Using Proposition 1
 we conclude that in this case a complex potential $V_1$
is
possible only for a complex value of $\a_2$.
The Hamiltonian
$h_1$ has thus the complex discrete level $E=\a_2$.
So, there are no ways to obtain a complex potential with a real
spectrum in this case.
The 2-SUSY transformation is reducible
if $u_1(x)=\psi_0(x)$ (ground state function)
and irreducible otherwise.

In case {\bf (c)} $u_1$ is still an eigenfunction of $h_0$,
$u_1(x)=\psi_k(x)$,
$\a_1=E_k$, $k=0,1,\ldots$
 and the level $E_k$ is not present in the spectrum
of $h_1$ but there are no restrictions on $u_2$.
Yet, the level $E=\a_2$ belongs to the spectrum of $h_1$ and
if we want for the Hamiltonian $h_1$ to have a real spectrum
 we have to choose $\a_2$ real. In this case complex potential
 differences can arise from formula (\ref{V1}) only if $u_2$ is a
 complex linear combination of two real linearly independent solutions
 of equation (\ref{1}) with $h=h_0$.
 As it is shown in \cite{MyLett1} if $\a_2=E_l$
 the  Hamiltonian $h_1$ becomes non-diagonalizable.
 The second order transformation
is irreducible provided $k>0$.
Another interesting feature we would like to
 mention is the possibility to get \pt-symmetric potentials by
appropriate choice of the function $u_2(x)$
and if $V_0(x)$ has this property.

\newcounter{exms}
\setcounter{exms}{1}

 {\it Example \arabic{exms}.}
 \addtocounter{exms}{1}
For  $x\in[-\pi,\pi]$
take $u_1=\sin(n_0 x)$,
($\a_1=n_0^2$).
 $n_0=1,2,\ldots$,
$u_2=\cos(ax+b)$
($\a_2=a^2$),
$a\in\Bbb R$,
$a\ne n_0$,
$\Im(b)\ne0$.
 The potential $V_1$ is given by
\be\label{ex0}
V_1=(n_0^2-a^2)\,
\frac{n_0^2[\cos(2ax+2b)+1]+a^2[\cos(2n_0x)-1]}%
{[n_0\cos(n_0x)\cos(ax+b)+a\sin(n_0x)\sin(ax+b)]^2}\,.
\ee
If $a\ne\pm n/2$, $n=1,2,\ldots$ the spectrum
of the Hamiltonian $h_1$ with potential
(\ref{ex0}) consists of
all levels of $h_0=-\p_x^2$
which are $E_{n-1}=n^2/4$
except for $E=n_0^2$
 and an additional level $E_a=a^2$.
If $\Re(b)=0$ this potential is explicitly \pt-symmetric.

One can find other examples
of potentials one can get under these conditions
 in \cite{MyLett1}.

 Consider finally the case {\bf (d)}.
Using Proposition 1
  we conclude that if both $\a_1$ and $\a_2$ are real there is no
 way to obtain a complex potential difference.
So, to be able to produce a complex potential $V_1$ we have to choose
 at least one of
 $\a$s (say $\a_1$) complex.
 In this case $u_1$
 is nodeless and the intermediate Hamiltonian $\wt h$ is well-defined
 in $L^2(a,b)$ but its spectrum is completely different from the
 spectrum of $h_0$ since
corresponding 1-SUSY transformation breaks the zero boundary
condition at $x=b$. We thus
   can construct an irreducible SUSY model
 of a new type.
Here also one can  get \pt-symmetric potentials
if
$V_0(x)$ is \pt-symmetric and
$\a_2=\a_1^*$.
This is
 readily seen for a symmetric interval $b=-a$. The last comment
 here is that according to (\ref{fial}) neither $\vfi_{\a_1}$ nor
 $\vfi_{\a_2}$ satisfy the zero boundary conditions. Therefore
 the Hamiltonian $h_1$ is strictly isospectral to $h_0$
 and, hence, its spectrum is purely real.

 {\it Example \arabic{exms}.}
 \addtocounter{exms}{1}
 For $x\in[-\pi,\pi]$ taking
 $u_1=\sin(a_1(x+\pi))$
($\a_1=a_1^2$)
 and $u_2=\sin(a_2(x-\pi))$,
 ($\a_2=a_2^2$),
 $a_1\ne a_2$,
 $\Im(a_1^2)\ne0$,  $\Im(a_2^2)\ne0$
 one gets the following potential:
 \be\label{ex1}
V_1=(a_2^2-a_1^2)
\frac{a_2^2[1-\cos(2a_1(x+\pi))]-
a_1^2[1-\cos(2a_2(x-\pi))]}%
{[a_1\cos(a_1(x+\pi))\sin(a_2(x-\pi))-
a_2\cos(a_2(x-\pi))\sin(a_1(x+\pi))]^2}
 \ee
which is  \pt-symmetric provided $a_2=a_1^*$.
The spectrum of the Hamiltonian $h_1$ coincides with the spectrum
of $h_0$, $E_{n-1}=n^2/4$, $n=1,2,\ldots$.
Although 1-SUSY transformations both with $u=u_1$ and
$u=u_2$
produce potentials regular in the interval $(-\pi,\pi)$ they
do not preserve the zero boundary conditions at both limiting
points of $(a,b)$ and therefore the 2-SUSY transformation is
irreducible.

 For the confluent case as it follows from (\ref{fiConfl}) the
 function $u(x)$ should vanish at either bound of the interval
 $[a,b]$ and therefore it is one of the eigenfunctions of $h_0$,
 $u=\psi_k$ and $\a=E_k$, $k=0,1,\ldots$.
 Therefore to obtain a complex
 potential $V_1$ (\ref{V1}) one has to choose the  constant
 $c$ in (\ref{WConfl}) complex.
 This transformation is irreducible provided $u\ne\psi_0$.
  It keeps the spectrum unchanged since the function
(\ref{fiConfl}}) satisfies the zero boundary conditions.
In some cases (i.e. $V_0(-x)=V_0(x)$, $b=-a$, $x_0=0$ and $\mbox{Re}(c)=0$)
the potential $V_1(x)$ is \pt-symmetric.

{\it Example \arabic{exms}}.
 \addtocounter{exms}{1}
 For $x\in[-\pi,\pi]$ taking
 $x_0=0$
we get the potentials
\be
V_1=\mp n_0^2\,\frac{n_0(2c+x)\sin(n_0x)+2\cos(n_0x)\pm 2}%
{[\sin(n_0x)\pm n_0(2c+x)]^2}\,.
 \ee
Here the upper sign corresponds to
 $u=\cos(n_0x/2)$, $n_0=3,5,7,\ldots$
 and the lower sign corresponds to
 $u=\sin(n_0x/2)$, $n_0=4,6,8,\ldots$,
 ($\a=n_0^2/4$),
 $\Im(c)\ne0$).
 For $\mbox{Re}(c)=0$ these potentials are \pt-symmetric.
The Hamiltonian $h_1$ is
isospectral with $h_0=-\p_x^2$.

\subsection{Scattering potentials on a semiaxis}

As it was already mentioned at the beginning of Section 3.2,
1-SUSY
transformation keeps unchanged the zero boundary condition
(at the origin in the current case)
 if $u_1(0)=0$, $u_1(x)\ne0$ $\forall x\in(0,\infty)$.
 The zero boundary condition at the infinity
for a transformed function
  is satisfied for any 1-SUSY transformation provided the
  transformation operator acts on a function
  vanishing at the infinity.
  Therefore any 2-SUSY transformation involving a
  transformation function vanishing at the origin and nodeless in
  the positive semiaxis is reducible.

Consider spectral problem (II).
To keep the boundary condition at the origin
after 2-SUSY transformation
according to
(\ref{fi1}) one has to impose the same condition on one of the
transformation functions, say $u_1(x)$,
i.e. $u_1(0)=0$.
For the second order transformation to be irreducible
 one has to take care of presence of a positive node in $u_1(x)$.
According to Proposition 1 if $\a_1$ is complex the function
 $u_1(x)$ is nodeless in $(0,\infty)$
 and 2-SUSY transformation is reducible.
 Therefore to construct an irreducible 2-SUSY transformation we
 have to choose only real values for $\a_1$.
 According to the second part of the same proposition
 the function $u_1(x)$ can be chosen real without any loss of
 generality.
So, we choose $\a_1$ real,
$u_1(x)$ real-valued
 and for
 $\a_2\ne\a_1$
we enumerate
  the following possible choices for $u_2$:\vspace{-.5em}
 \begin{description}
\item[(a)] $u_2(0)=0$;
\item[(b)] $u_2(0)\ne 0$ and $u_2(\infty)=0$;
\item[(c)] $u_2(0)\ne0$ and $u_2(\infty)=\infty$;
\item[(d)] $u_2(0)\ne0$ and $u_2(x)$ has an oscillating asymptotics at
the infinity.
\end{description}

For all cases (a)-(d) if $\a_1$ is a point of the discrete spectrum of $h_0$,
the function $u_1$ is a
eigenfunction
 of the problem (I) (it is a bound state) and the
point $E=\a_1$ does not belong to the discrete
spectrum of the problem (II).

In case {\bf (a)}
by the same reason as it was explained above
an irreducible 2-SUSY transformation is possible if
$\a_2$ is real but
according to Proposition 1
it may produce only a real potential $V_1$.
Nevertheless, one can get interesting complex potentials by
reducible transformations with a complex $\a_2$.
If $\a_1$ does not belong to the discrete spectrum of $h_0$
(i.e. $|u_1(\infty)|=\infty$) the 2-SUSY transformation is
isospectral.

 {\it Example \arabic{exms}.}\addtocounter{exms}1
Choose $u_1=\sin(k_0x)$ $(\a_1=k_0^2>0)$
and $u_2=\sinh(ax)$ $(\a_2=a^2\in\Bbb C,\ {\rm Im}\a_2\ne0)$.
Formula (\ref{V1}) gives the potential
\be\label{ex3}
V_1=(k_0^2+a^2)\,
\frac{k_0^2[\cosh(2ax)-1]+a^2[\cos(2k_0x)-1]}%
{[k_0\cos(k_0x)\sinh(ax)-a\sin(k_0x)\cosh(ax)]^2}\,.
\ee
 Formula (\ref{fial}) gives an oscillating solution of the
\Sc equation with the potential (\ref{ex3})
at $E=k_0^2$ but it is irregular at the
origin.
A regular at the origin solution at $E=k_0^2$ is unbounded as
$x\to\infty$.

Similar to the case (a) in case {\bf (b)}
with a real $\a_2$ according to Proposition 1
 the potential $V_1$ remains real.
 Therefore to obtain  a
complex $V_1$ one has to choose
 $\a_2$ complex.
  The function $\vfi_{\a_2}(x)$ (\ref{fial})
 is vanishing at the origin but
increasing at the infinity.
So, the 2-SUSY transformation
creates a complex potential
with a real spectrum
coinciding with the spectrum of $h_0$ with the possible exception
of the point $E=\a_1$
(if it belongs to the spectrum of $h_1$).
Here 2-SUSY transformation is reducible only if $u_1(x)$ is
nodeless in $(0,\infty)$.
According to Proposition 1 the function
$u_2$ with a complex $\a_2$
is nodeless
and
produces a ``good"
intermediate potential but the
first order
transformation operator
based on $u_2$
does not
transform eigenfunctions of $h_0$ into eigenfunctions of $\wt h$.
So, the 2-SUSY transformation is irreducible if $\a_1$ is real,
the function
$u_1(x)$ has a node in $(0,\infty)$ and $\a_2$ is complex.

 {\it Example \arabic{exms}.}\addtocounter{exms}1
Once again we choose $u_1=\sin(k_0x)$ ($\a_1=k_0^2>0$)
but $u_2=e^{ax}$
($\a_2=-a^2$, $\mbox{Re}(a)<0$, $\Im(a^2)\ne0$). The potential $V_1$ reads
\be\label{ex4}
V_1=\frac{2k_0^2(k_0^2+a^2)}{[k_0\cos(k_0x)-a\sin(k_0x)]^2}\,.
\ee

In contrast to the case (b) in case {\bf(c)} function
$\vfi_{\a_2}$  (\ref{fial}) is an eigenfunction of $h_1$ and
$E=\a_2$ is its spectral point.
Therefore to get  Hamiltonian $h_1$ with a real spectrum we
have to choose real values for $\a_2$.
 Complex potentials can arise in this case if
 the function
$u_2$ is a complex linear combination of two real
linearly independent solutions to equation (\ref{1}) at $E=\a_2$.
So,
the 2-SUSY transformation creates a new energy level $E=\a_2$.
It is reducible if $u_1(x)$ is nodeless in $(0,\infty)$ and
irreducible otherwise.

 {\it Example \arabic{exms}.}\addtocounter{exms}1
 Choice $u_1=\sin(k_0x)$
($\a_1=k_0^2$)
 and $u_2=\cosh(ax+c)$ ($\a_2=-a^2$),
$k_0,a\in \Bbb R$ and $\mbox{Im}(c)\ne0$
 results in
the potential
\be
V_1=(k_0^2+a^2)\,
\frac{a^2[1-\cos(2k_0x)]+k_0^2[1+\cosh(2ax+2c)]}%
{[k_0\cos(k_0x)\cosh(ax+c)-a\sin(k_0x)\sinh(ax+c)]^2}
\ee
having a discrete level $E=-a^2$.

In case {\bf(d)} $\a_2>0$.
Therefore
like in the previous case
 to have a complex $V_1$
  $u_2$ should be a complex linear
combination of two linearly independent solutions to equation
(\ref{1}). The function $\psi_{\a_2}$ belongs to the continuous
spectrum of $h_1$ which has a purely real spectrum.
In some cases indicated in \cite{MyLett2}
the point $E=\a_2$ is a spectral singularity of the
Hamiltonian $h_1$.

 {\it Example \arabic{exms}.}\addtocounter{exms}1
To illustrate this case we
take $u_1=\sin(k_0x)$ ($\a_1=k_0^2$) and
$u_2=\sin(k_1x+c)$ ($\a_2=k_1^2$),
$k_0,k_1\in \Bbb R$, $k_0\ne k_1$, $\Im(c)\ne0$ thus getting the
potential
\be
V_1=(k_1^2-k_0^2)\,
\frac{k_1^2[1-\cos(2k_0x)]-k_0^2[1-\cos(2k_1+2c)]}%
{[k_0\cos(k_0x)\sin(k_1x+c)-k_1\sin(k_0x)\cos(k_1x+c)]^2}\,.
\ee

For the confluent transformation
(\ref{V1}), (\ref{WConfl}), (\ref{fiConfl})
to keep the zero boundary conditions we have to choose
 $u(0)=0$.
 Therefore by the same reason as it was explained above to get an
 irreducible 2-SUSY transformation we have to choose
 real values for $\a$
 leading to a real-valued function $u(x)$.
To obtain a complex potential $V_1$
one has to choose for the
constant $c$ from (\ref{WConfl}) a complex value.
If $u(x)$ decreases at the infinity
(i.e. it is an eigenfunction of $h_1$)
the function $W_c(x)$ (\ref{WConfl}) is finite at the infinity and
$\vfi_\a$ (\ref{fiConfl}) is an eigenfunction of $h_1$.
For $x_0$ one can choose both $x_0=0$ and $x_0=\infty$.
Since
$u(x)$ is square integrable
different choices for $x_0$
 affect only the value of $c$.
If $u(x)$
 increases at the infinity,
one can choose $x_0=0$.
Therefore the function $W_c(x)$ increases as
$u^2(x)$ when $x\to\infty$ and the function
$\vfi_\a$ (\ref{fiConfl}) is an eigenfunction of $h_1$ too.
 If
$\a>0$ the function $u(x)$ oscillates at the infinity as a linear
combination $c_1\exp(-i\sqrt \a x)+c_2\exp(i\sqrt\a x)$. The
integrand in (\ref{WConfl}) increases as a linear function of $x$
and the potential $V_1$ keeping its oscillating behavior
decreases like $1/x^2$. Therefore it does not satisfy
condition (\ref{cond1}) and it is not a scattering potential.
This
leads to the existence of a discrete level embedded into the
continuous spectrum since the function $\vfi_\a$ (\ref{fiConfl})
is vanishing at the origin and square integrable for $x\in [0,\infty)$.
So, to get a complex potential by confluent transformation
(\ref{V1}), (\ref{WConfl}) one has to choose $\a$ real,
$c$ complex
and
$x_0=0$.
If $\a$ is not a discrete spectrum level then it appears as a new
energy level for the Hamiltonian $h_1$. If $u(x)$ is nodeless in
$(0,\infty)$ the transformation is reducible and irreducible
otherwise.

 {\it Example \arabic{exms}.}\addtocounter{exms}1
We choose $u=\sin(k_0x)$ ($\a=k_0^2>0$)
and replace $c\to c/2$ ($\Im(c)\ne0$).
The potential
\be
V_1=32k_0^2\sin(k_0x)\,\frac{\sin(k_0x)-k_0(x+c)\cos(k_0x)}%
{[\sin(2k_0x)-2k_0(x+c)]^2}
\ee
has the discrete level $E=k_0^2$ embedded into continuum of the
scattering states.

\subsection{Scattering and confining potentials on the whole real
line}

For a scattering potential the logarithmic derivative of any
decreasing or increasing at $x\to\pm\infty$
solution $u(x)$ of the \Sc equation is asymptotically constant. For a
confining potential
the logarithmic derivative of any similar solution $u(x)$ is usually such
that the product $\psi_E(x)u'(x)/u(x)$ increases or decreases
together with $\psi_E(x)$
so that in the last case its asymptotics is square integrable;
the behavior we will assume to take place.
For a scattering potential if $u(x)$ has an oscillating
asymptotics and $\psi_E(x)$ has an exponentially decreasing one
the product  $\psi_E(x)u'(x)/u(x)$ is exponentially
decreasing also so that $\wt L\psi_E$ belongs to the discrete spectrum
of $\wt h$ provided $\psi_E$ belongs to the discrete
spectrum of $h_0$.
This means that if $u_1(x)$ is nodeless in $\Bbb R$ it produces a
``good" intermediate Hamiltonian $\wt h$.
Thus, if $\a$ is real according to Proposition 1
any essentially complex-valued
solution of equation (\ref{1})
(i.e. a solution $u(x)$ such that $u'(x)/u(x)$ is a
complex-valued function)
is suitable for getting
a ``good" complex first order potential difference.
Therefore if both $\a_1$ and $\a_2$ are real,
any 2-SUSY transformation
which may produce a complex potential $V_1$
is reducible.
Let at least one of the factorization constants, say
$\a_1$ be complex (another constant, $\a_2$, may be both real and complex).
If
 $u_1(x)$ vanishes at one of the infinities then
according to Proposition 1 it
does not have real nodes and 2-SUSY transformation is reducible
also.
If  $|u_1(x)|\to\infty$ when $|x|\to \infty$
then according to (\ref{fial}) the potential
$V_1$ has a complex eigenvalue $E=\a_1$.
We conclude, hence, that no irreducible 2-SUSY transformations
giving a complex potential $V_1$ with a real spectrum exist.
Of course this does not mean that such transformations cannot
create complex potentials with a real spectrum but this means that
any such a transformation can always be presented as a chain of
two `good' 1-SUSY transformations (with a possibility for the
intermediate potential to be complex).
In particular, if
$\Im(\a_{1,2})\ne0$  and
the functions
$u_1(x)$, $u_2(x)$ vanish  at different infinities
(i.e. for a scattering potential they are two Jost solutions)
one can get a complex
 potential isospectral with $h_0$
(hence, its spectrum is purely real).
If
for $V_0(x)=V_0(-x)$
 in addition
 $\a_2=\a_1^*$, $V_1(x)$ is \pt-symmetric.
The last comment here is that irreducible 2-SUSY transformations
can produce potentials with complex eigenvalues.

{\it Example \arabic{exms}.}\addtocounter{exms}1
Take $u_1=\sinh(a_1(x-x_1))$ ($\a_1=-a_1^2$),
$u_2=\sinh(a_2(x-x_2))$  ($\a_2=-a_2^2$),
$\Im(a^2_1)\ne0$, $\Im(a^2_2)\ne0$, $a_1\ne a_2$,
$x_1,x_2\in \Bbb R$.
The potential
\be
V_1=(a_2^2-a_1^2)\,
\frac{a_2^2[1-\cosh(2a_1(x-x_1))]-a_1^2[1-\cosh(2a_2(x-x_2)]}%
{[a_2\cosh(a_2(x-x_2))\sinh(a_1(x-x_1))-
a_1\cosh(a_1(x-x_1))\sinh(a_2(x-x_2))]^2}
\ee
has discrete levels at $E=-a_1^2$ and $E=-a_2^2$.
If $x_2=-x_1$ and $a_2=a_1^*$ it is explicitly
\pt-symmetric.

Consider finally the confluent case. If $\Im(\a)\ne0$ and $u(x)$
increases at both infinities, as it follows from (\ref{fiConfl})
  the point $E=\a$ belongs to the spectrum of $h_1$. If $u(x)$
  decreases at one of the infinities then according to Proposition
  1 the function $u(x)$ has no real nodes and 2-SUSY
  transformation is reducible. Hence, irreducible 2-SUSY
  transformations
creating complex potentials $V_1$ with a real spectrum
   are possible only with real values of $\a$ and $\Im(c)\ne0$.
In this case the function (\ref{fiConfl}) has a decreasing
asymptotical behavior both for increasing and decreasing $u(x)$
as well as for $u(x)$ having an oscillating asymptotical behavior.
Therefore  the level $E=\a$ belongs to the discrete spectrum of $h_1$.
Simple potentials one can get in this way correspond to
the choice of an eigenfunction
(e.g. a bound state function) of $h_0$
 as the transformation function
 $u(x)$
 since in many cases it is described
in terms of elementary functions.
 We do not illustrate these possibilities by examples and refer an
 interested reader to existing literature where this is done
  \cite{Milanovich}.

\section{Concluding remarks and some perspectives}

In summary,
 in this Letter
a careful analysis
aimed to distinguish irreducible transformations between all
2-SUSY transformations for
three important Sturm-Liouville problems, namely, regular problem,
problem on a half axis and problem on the whole real line is
given.
We remind that we call irreducible those
second order transformations for which
 either (i)
the intermediate Hamiltonian of corresponding chain of two
transformations is not well defined in the same Hilbert
space as the initial and final Hamiltonians
or (ii)
the intermediate Hamiltonian is well defined but its
eigenfunctions cannot be obtained by acting with the
intertwining operator either on eigenfunctions of the
initial Hamiltonian or on those of the final Hamiltonian.
It is shown that for the whole real line the only
possibility for such a transformation to be irreducible
corresponds to the confluent case, i.e.
to a chain of transformations
with coinciding factorization constants.
For problems on a half line and
on a finite interval there are more possibilities.
In particular,
 transformations of type (ii)
  lead to new irreducible SUSY models.

Using the property of SUSY transformations to provide us with a
general solution of the \Sc equation at any fixed value of the
energy one can observe
in examples 4-7
an unusual property of SUSY transformations and
an intriguing phenomenon concerning spectral properties of
non-Hermitian operators.
 In the usual practice of SUSY transformations
\cite{JPA,revs,BS} if a transformation function corresponds to a
spectral point (for the usual non-confluent
 case this may be only a point of the
discrete spectrum) this point is deleted by the transformation.
In examples 4-7 we used a transformation function corresponding
to a point in
the continuous part of the spectrum
of the initial Hamiltonian
but in contrast to
conventional SUSY transformations
(i.e. transformations between Hermitian operators)
 this point now still belongs to
the continuous spectrum of the transformed problem.
This statement follows from a general spectral theorem
(see e.g. \cite{Glazman}) according to which the spectrum
of a closed operator is a closed
set and the property that if a point is removed from a closed
interval of the real axis it is transformed into two (semi-)open
subintervals.
Of course, the operator $h_1$ with the initial domain of
definition as described in Section 2 is not closed but it is
closable and its closure coincides with $h_1^{++}$
(see e.g. \cite{Plesner}, theorem 6.3.2).
Since the eigenfunctions of the operator $h_1^+$ coincide with
complex conjugate eigenfunctions of $h_1$, the operator $h_1^{++}$ has
the same system of eigenfunctions as $h_1$.
Moreover, the analysis of solutions of the \Sc equation
at $E=k_0^2>0$
shows that a solution vanishing at the origin is unbounded.
This contrasts with the usual quantum mechanical requirement
that continuous spectrum eigenfunctions should be bounded.
 We think that instead of imposing the above quantum mechanical
 requirement on continuous spectrum eigenfunctions one should use
 the fact that these functions being ordinary locally integrable
 functions are in fact generalized eigenfunctions of the
 hamiltonian $h_1$ and should be considered as functionals over
 the domain of definition of $h_1$.
(In other words one has to involve the notion of the Gelfand
triplet into analysis \cite{Gelfand}.)
From this point of view the question whether
continuous spectrum eigenfunctions are bounded
 or not has no importance.
 Nevertheless, if necessary one can analyze the growth of
 generalized eigenfunctions for $x\to\infty$
 (see theorem 6 in section 55 of \cite{Glazman})
 as they are ordinary locally integrable functions.

We think that the results of the present Letter are important in
view of the notion of $\cN$-fold supersymmetry \cite{NSYM}.
As far as this notion is applied to Hamiltonians of type
(\ref{h01})
with supercharges built of differential intertwining operators
acting in the same Hilbert space as the components of the
super-Hamiltonian
one can always apply a theorem \cite{BS} to factorize an $N$th
order in derivative intertwining operator to a superposition of
 first order operators thus replacing an $N$th order
 transformation by a chain of only first order transformations.
 If
 all intermediate Hamiltonians of the chain are defined in the
 same Hilbert space as the initial and final Hamiltonians and at
 any step of transformations the eigenfunctions of two neighbor
 Hamiltonians are connected by corresponding transformation
 (i.e. intertwining)
 operator the $\cN$-fold supersymmetry is reducible and,
  actually,
 all properties of the final Hamiltonian
(as well as any intermediate Hamiltonian)
 can be understood at the
 level of a chain of simpler first order (i.e. usual)
 supersymmetry transformations. Evidently this is not the case if at least one
 Hamiltonian of the chain is not well-defined
(case (i) of irreducible supersymmetry)
  or at least for one of the
 Hamiltonians we will not be able to get eigenfunctions by applying
 transformation operator to eigenfunctions of its SUSY partner
 (case (ii) of irreducible supersymmetry).
The main result of the present Letter consists in formulating conditions
 on transformation functions
 to produce the simplest irreducible
two-fold supersymmetry
between real and complex potentials.

Another field of application of our results is the supersymmetric
approach to the inverse scattering problem
(so called supersymmetric inversion \cite{inver}).
Since irreducible chains of transformations proved to be very
efficient in supersymmetric inversion for usual Hermitian
operators \cite{irCh} we hope that our results open a way for
wider
applications of the method of supersymmetric inversion to complex
potentials previously used for obtaining complex optical potentials
 only at the level of reducible transformations
\cite{SBc}.

\section*{Acknowledgments}

The work is partially supported by
the grants SS-5103.2006.2 and RFBR-06-02-16719.
The author is grateful to M.V. Ioffe for pointing out some
references and to anonymous referee for useful comments.


\begin{thebibliography}{99}


\bibitem{Pauli}
P.A.M. Dirac, Proc. Roy. Soc. London {A 180} (1942) 1;\\
W. Pauli, {Rev. Mod. Phys.} {15} (1943) 175;\\
T. D. Lee, Phys. Rev. {95} (1954) 1329;\\
T. D. Lee and G.C. Wick, Nucl. Phys. B 9 (1969) 209.

\bibitem{Gupta}
S.N. Gupta, Phys. Rev. {77} (1950) 294;\\
K. Bleuler, Helv. Phys. Acta. { 23} (1950) 567.\\


\bibitem{Bender}C.M. Bender and S. Boettcher,
{Phys. Rev. Lett.} {80} (1998) 5243.

\bibitem{BBJ}C.M. Bender, D.C. Brody and H.F. Jones,
 Phys. Rev. Lett. { 89} (2002) 270401;\\
C.M. Bender, D.C. Brody and H.F. Jones,
{Phys. Rev. Lett.} { 92} (2004) 119902 (erratum).

\bibitem{BCM}C.M. Bender, J.-H. Chen, and K.A. Milton,
{J. Phys. A: Math. Gen.} {39} (2006) 1657.

\bibitem{ChJP}
M. Znojil, Ed. {Pseudo-Hermitian Hamiltonians
in Quantum Physics} (special issue of Czech. J. Phys.
55 (2005) No 9).

\bibitem{QuasiH}F.G. Scholtz, H.B. Geyer and F.J.W. Hahne,
Ann. Phys. 213 (1992) 74.

\bibitem{SS}L. Solombrino, J. Math. Phys. 43 (2002) 5439;\\
G. Scolarici and L. Solombrino, J. Math. Phys. 44 (2003) 4450.


\bibitem{Mostafa}A. Mostafazadeh, J. Math. Phys. 43 (2002) 3944.

\bibitem{MostafaP}A. Mostafazadeh, Preprint quant-ph/0603059.

\bibitem{Naimark}
M. A. Naimark,  Linear differential operators,
Nauka, Moscow, 1969.

\bibitem{Z2}
M. Znojil, F. Cannata, B. Bagchi and R. Roychoudhury,
{Phys. Lett. B} {483} (2000) 284;\\
M. Znojil 2001 Czech. J. Phys. {51} (2001) 420.


\bibitem{JPA}
I. Aref'eva, D.J. Fern{\'a}ndez, V. Hussin, J. Negro,
L.M. Nieto, and B.F. Samsonov, eds., Progress in Supersym-
metric Quantum Mechanics (special issue of J. Phys. A:
Math. Gen.  {37} (2004) No 43).

\bibitem{MyLett1}B.F. Samsonov,
{J. Phys. A: Math. Gen.} 38 (2005) L397.

\bibitem{MyLett2}B.F. Samsonov,
{J. Phys. A: Math. Gen.} 38 (2005) L571.

\bibitem{Cannata1}
A.A. Andrianov, F. Cannata, J.-P. Dedonder and M.V. Ioffe,
{ Int. J. Mod. Phys. A} { 14} (1999) 2688.


\bibitem{Cannata2}
F. Cannata, G. Junker and J. Trost,
{ Phys. Lett. A} { 246} (1998) 219.

\bibitem{revs}B. Mielnik and O. Rosas-Ortiz,
{J. Phys.} A: Math. Gen. {43} (2004) 10007;\\
D.J. Fernandez, {AIP Conf. Proc.} { 744} (2005) 236.

\bibitem{Milanovich}V. Milanovi\'c and Z. Ikoni\'{c},
{Phys. Lett. A} { 293} (2002) 29;\\
J.S. Petrovi\'{c}, V. Milanovi\'{c} and Z. Ikoni\'{c},
{ Phys. Lett. A} { 300} (2002) 595.



\bibitem{Andr2}A.A. Andrianov, M.V.  Ioffe and V.P. Spiridonov,
{Phys. Lett.} { A 174} (1993) 273;\\
A.A. Andrianov, F. Cannata, J.P. Dedonder and M.V. Ioffe,
{ Int. J. Mod. Phys.} {A 10} (1995) 2683.


\bibitem{S}B.F. Samsonov, { Mod. Phys. Lett.} { 11} (1996)
1563;\\
B.F. Samsonov, { Phys. Lett.} { A 263} (1999) 274.

\bibitem{AndrCan}A. A. Andrianov and F. Cannata,
{J. Phys. A: Math. Gen.} {A 37} (2004) 10297.

\bibitem{SP}B.F. Samsonov and A. Pupasov,
 Russ. Phys. J. 48(10), (2005) 20.


\bibitem{BS}V.G. Bagrov and B.F. Samsonov,
{Theor. Math. Phys.} {104} (1995) 356;\\
V.G. Bagrov and B.F. Samsonov, {Phys. Part. Nucl} {28} (1997)
374.

\bibitem{Krein}M. Crum, {Quart. J. Math.} {6} (1955) 263;\\
M.G. Krein, {DAN SSSR}
(Doklady Akademii Nauk SSSR) {113} (1957) 970.

\bibitem{Berezin}F.A. Berezin M.A. and M.A. Shubin,
{The \Sc equation}, Kluwer, Dordrecht, 1991.

\bibitem{Glazman}I.M. Glazman, Direct methods of qualitative
spectral analysis of singular differential operators,
Moscow, Fizmatgiz, 1963.

\bibitem{Plesner}A.I. Plesner, Spectral theory of linear
operators, Nauka, Moscow, 1965.

\bibitem{Gelfand}I.M. Gelfand and G.E. Shilov,
Generalized functions, vol. 1-3,
Academic, New York, 1964;\\
I.M. Gelfand and N.Ya. Vilenkin,
Generalized functions, vol. 4,
Academic, New York, 1964.

\bibitem{NSYM}H. Aoyama, M. Sato and T. Tanaka,
Nucl. Phys. B 619 (2001) 105;\\
H. Aoyama, N. Nakayama, M. Sato and T. Tanaka,
Phys. Lett. B 519 (2001) 260;\\
T. Tanaka, Nucl. Phys. B 662 (2003) 413;\\
A. Gonzales-Lopes and T. Tanaka,
J. Phys. A: Math. Gen. 39 (2006) 3715

\bibitem{inver}D. Baye, Phys. Rev. Lett. 58 (1987) 2738;\\
D. Baye and J.-M. Sparenberg, Phys. Rev. Lett. 73 (1994)
2789;\\
J.-M. Sparenberg and D. Baye, Phys. Rev. C 55 (1997) 2175;\\
J.-M. Sparenberg, D. Baye, and H. Leeb, Phys. Rev. C 61
(2000) 024605.\\
B. F. Samsonov and F. Stancu, B.F.,
Phys. Rev. C 66 (2002) 034001.

\bibitem{irCh}B. F. Samsonov and F. Stancu,
Phys. Rev. C 67 (2003) 054005;\\
 J.-M. Sparenberg, B. F. Samsonov, F. Foucart
 and D. Baye, preprint quant-ph/0601101.

\bibitem{SBc}J.-M. Sparenberg and D. Baye,
Phys. Rev. C 54 (1996) 1309.

\end{thebibliography}
\end{document}